\documentclass[aps,showkeys,superscriptaddress,preprint,tightenlines]{revtex4}
\usepackage{hyperref}
\usepackage{times}
\usepackage{amsmath}
\usepackage{graphicx}
\usepackage{epsfig}
\usepackage{dcolumn}
\usepackage{bm}
\usepackage{color}
\usepackage{longtable}
\usepackage{array}
\usepackage{multirow}
\usepackage{subfigure}

\begin{document}

\title{Tests of CP symmetry in entangled hyperon anti-hyperon pairs at \mbox{BESIII}}

\author{Wenjing Zheng}
 \email{zhengwenjing@ihep.ac.cn}
 \affiliation{Institute of High Energy Physics, Chinese Academy of Sciences, Beijing 100049, China}
 \affiliation{University of Chinese Academy of Sciences, Beijing 100049, China}

 \author{Andrzej Kupsc}
 \email{Andrzej.Kupsc@physics.uu.se}
 \affiliation{Department of Physics and Astronomy,
Uppsala University, Box 516, SE-75120 Uppsala, Sweden}
 \affiliation{National Centre for Nuclear Research,
Pasteura 7, 02-093 Warsaw, Poland}

 \author{Simone Pacetti}
 \email{simone.pacetti@pg.infn.it}
 \affiliation{INFN Sezione di Perugia, 06100 Perugia, Italy}
 \affiliation{Università di Perugia, 06100 Perugia, Italy}

\author{Francesco Rosini}%
\email{francesco.rosini@pg.infn.it}
 \affiliation{INFN Sezione di Perugia, 06100 Perugia, Italy}
 \affiliation{Università di Perugia, 06100 Perugia, Italy}

 \author{Nora Salone}
\email{nora.salone@ncbj.gov.pl}
\affiliation{National Centre for Nuclear Research,  Pasteura 7, 02-093 Warsaw, Poland}

 \author{Xiongfei Wang}
 \email{wangxiongfei@lzu.edu.cn}
 \affiliation{Lanzhou University,  Lanzhou 730000,  China}
 
\author{Shuang-shi Fang}
 \affiliation{Institute of High Energy Physics, Chinese Academy of Sciences,
 Beijing 100049, China}
 \affiliation{University of Chinese Academy of Sciences, Beijing 100049, China}
  \affiliation{Institute of Physics, Henan Academy of Sciences, Zhengzhou, 450046, China}
\begin{abstract}

Decays of charmonium into hyperon and antihyperon pairs provide a pristine laboratory for exploring hyperon properties, such as their polarization and decay parameters, and for conducting tests of fundamental symmetries. This brief review highlights the significant progress made in precise tests of CP symmetry at BESIII using entangled hyperon-antihyperon pairs, including  $\Lambda\bar{\Lambda}$, $\Sigma\bar{\Sigma}$, $\Xi\bar{\Xi}$ and $\Lambda\bar{\Sigma}$, selected from the high statistics of $J/\psi$ and $\psi(3686)$ events produced in $e^+e^-$ annihilations. These recent findings have sparked renewed interest in both theoretical and experimental aspects of hyperon physics, but there is still much room for improvement to reach the Standard Model expectations. To address this challenge, the prospects for future investigations on CP asymmetry at next-generation experiments are discussed.

\end{abstract}


\maketitle

\section{Introduction}
\label{Sec:intro}

At the particle physics level, one would expect that matter and antimatter should exist symmetrically in the universe. However, the observable natural world provides no evidence of primordial antimatter. One possible explanation requires violation of CP and P symmetries, suggesting that particles and their antiparticles exhibit slight differences in interactions, resulting in an evolution that favors a universe dominated by matter. The discovery of parity violation in weak interactions in 1956 marked a significant milestone. Subsequently, CP violation has been observed in processes involving neutral strange (K), beauty (B), and charm (D) mesons. However, to date, no CP violation has been detected in hyperon decays, presenting a promising and significant avenue for exploring CP violation.

Hyperons can be readily produced in colliders, but generating the necessary quantity as required by the expected CP violation signal in the Standard Model (SM) remains a formidable challenge. The initial endeavor was the HyperCP experiment, which used spin unpolarized  $\Xi$ hyperons produced from a fixed target setup. Through the analysis of the decay $\Xi^-\rightarrow\Lambda\pi^-$, the experimental findings revealed no discernible CP asymmetry. While this experiment enhanced precision by a factor of 20 compared to previous studies, it still fell short by two orders of magnitude in achieving the SM predictions.

In recent years, the Beijing Spectrometer III (BESIII) experiment~\cite{BESIII:2009fln} has implemented an innovative and highly sensitive experimental approach that leverages the quantum correlations of hyperons and cascade decays to investigate the asymmetry between matter and antimatter. This marks a significant advancement in hyperon CP violation research, garnering considerable attention from international colleagues in the field. The BESIII experiment serves as a crucial facility for conducting extensive and impactful research in  $\tau$-charm physics. Since its establishment in 2008, BESIII has collected the world's largest datasets at the peaks of charmonium states, $e.g.$ 10 billion $J/\psi$ events~\cite{BESIII:2021cxx} and 2.7 billion $\psi(3686)$events~\cite{BESIII:2024lks}, produced from the $e^+e^-$ annihilations.  
 The abundant production of hyperons in  $J/\psi$ and $\psi(3686)$decays provides a clean laboratory for investigating hyperon properties, such as their polarization and decay parameters, and conducting tests of fundamental symmetries.

\section{Phenomenology of hyperon decays}

The phenomenology of CP violation in hyperon decays can be found in several excellent references~\cite{Lee:1957qs,Lee:1958qu,Faldt:2017kgy,Perotti:2018wxm}.
The most accessible signature for CP violation in spin-1/2 hperon decays, such as $\Lambda\rightarrow p\pi$, $\Sigma\rightarrow p\pi $ and $\Xi\rightarrow\Lambda\pi$ , is the comparison of the angular decay distributions of the daughter baryon with that of the anti-baryon in the conjugate decay. These distributions are not isotropic due to parity violation and are given by:

\begin{equation}
   \frac{dN}{d\cos\theta} = \frac{N_0}{2}(1 + \alpha{P_Y}\cos\theta),
\label{eq:1}
\end{equation}
where $P_Y$ is the parent hyperon polarization, $\theta$ is the daughter
baryon direction in the rest frame of the parent $Y$,
and $\alpha = 2\mbox{Re}(S^{\ast}P)/(|S|^2 + |P|^2)$
where $S$ and $P$ are the usual angular momentum amplitudes. Both amplitudes are characterized by the real part and weak and strong phases e.g. $S=|S|\exp(i\delta_S+i\xi^S)$.
In case of  CP conservation,  $\bar{\alpha}$ should be exactly the same as $-\alpha$; hence a difference in the magnitudes
of the hyperon and antihyperon alpha parameters,  $A_{CP} \equiv ({\alpha + \overline{\alpha})(} {\alpha - \overline{\alpha}})$, is an evidence of CP violation.   
Weak transitions can be classified by the amount of isospin changed $|\Delta I|$ between initial and final state which could be $1/2$, $3/2$ or $5/2$ and by the isospin of the final state. Here we limit to the transitions  $|\Delta I|=1/2$ since they dominate. Therefore, we can label phases of both the weak transitions $\xi$ and the final state strong rescattering $\delta$ by the isospin value of the final state.  For $\Lambda$ decays the  $A_{CP}$ asymmetry can be written as $A_{CP}^\Lambda=\tan(\delta_{\frac12}^P-\delta_{\frac12}^S)(\xi_{\frac12}^P-\xi_{\frac12}^S)$ 
 Numerically, 
using strong phases from~\cite{Hoferichter:2015hva,Salone:2022lpt}
$A_{CP}^\Lambda=-0.177(\xi_{\frac12}^P-\xi_{\frac12}^S)$. For $\Sigma^+\to p\pi$ both isospin $1/2$ and $3/2$ is reachable via $|\Delta I|=1/2$ transition and $A_{CP}^\Sigma=-0.066(\xi_{\frac12}^P-\xi_{\frac12}^S)+ 0.021(\xi_{\frac12}^S-\xi_{\frac32}^S)$. 

For $\Xi$ decays  $A_{CP}^\Xi=\tan(\delta_1^P-\delta_1^S)(\xi_{1}^P-\xi_{1}^S)$ but $\Lambda-\pi$ strong interaction is not well known. Instead, one can measure independently imaginary part of the interference term given by $\beta=2\mbox{Im}(S^{\ast}P)/(|S|^2 + |P|^2)$, via rotation angle of the spin vector.  This allows not only to determine $\delta_1^P-\delta_1^S$  but also test CP using  complementary asymmetry $B_{CP}^\Xi\equiv(\beta+\bar\beta)/(\alpha-\bar\alpha)$.  It is given  $B_{CP}^\Xi=\cot{}(\delta_1^P-\delta_1^S)(\xi_1^P-\xi_1^S)$ and ,since $\delta_1^P-\delta_1^S$ is small, it provides much improved sensitivity.
 
Within the framework of the SM, 
the CP asymmetries in $\Lambda$ and $\Xi$ hyperons are predicted to be in the ranges of $-3\times10^{-5}  \le A_{CP}^\Lambda \le 3\times10^{-5}$ and $0.5\times10^{-5}  \le A_{CP}^\Xi \le 6\times10^{-5}$\cite{Salone:2022lpt}, respectively. With the heavy-baryon chiral perturbation theory,   $A_{CP}^\Sigma$ was roughly estimated to be in the range of $(0.036-3.9)\times 10^{-4}$~\cite{Tandean:2002vy}.
Given presence of the physics beyond the SM, {\it e.g.},  the chromomagnetic-penguin interactions, the CP asymmetries in $\Lambda$ and $\Xi$ may enhance about one order of magnitude of  $-4.9\times10^{-4}  \le A_{CP}^\Lambda \le 5.9\times10^{-4}$ and $7\times10^{-4}  \le A_{CP}^\Xi \le 7\times10^{-4}$~~\cite{He:2022xra}.

\section{Experimental results at BESIII}

 As first proposed in Ref.~\cite{Faldt:2017kgy}, hyperon--antihyperon pairs form charmonium decays can be polarized 
if the state is produced in electron-positron annihilation with unpolarized beams since the photon, and therefore the charmonium, can only have $\pm 1$ helicities. 
At BESIII  the hyperons are produced in pairs  in  $e^+e^-$ collisions in reactions with sequential decays such as
$e^+e^-\to \psi\to Y\bar{Y}$,   ($\psi$ here denotes either the $J/\psi$ or the $\psi(3686)$) 
where the weak decays of the hyperons analyze their polarization. The combined angular distribution for a process like this one allows BESIII to simultaneously extract the parameters.

Taking advantage of the large $J/\psi$ and $\psi(3686)$ data samples and  the excellent performance of the detector,  BESIII  begun to challenge the measurement of CP violation effects in hyperons by reporting  a series of observations of hyperon polarization in charmonium decays which inspired both theoretical and experimental interests in hyperon physics.

\subsection{$J/\psi\rightarrow\Lambda\bar{\Lambda}$ }

\begin{figure}[!htbp]
    \centering
     \subfigure{
     \label{Lambda_mu}
	    \includegraphics[width=0.4\textwidth]{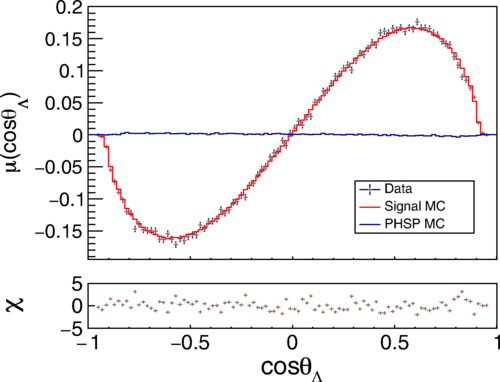}
     \put(-143,125){(a)}
     }\hspace{0.08\textwidth}
     \subfigure{
     \label{Lambda_alpha}
	    \includegraphics[width=0.4\textwidth]{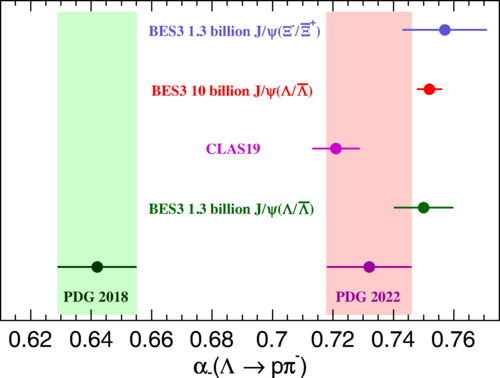}
     \put(-183,125){(b)}
     }
\caption{(a) Transverse polarization of $\Lambda$ hyperon as a function of the emission angle. (b) Results of the $\Lambda$ decay parameter from different experiments. The green band represents the Particle Data Group (PDG) 2018 value, and the pink band represents the PDG 2022 value.}\label{Lambda_polarization}
\end{figure}

In 2019, BESIII observed the hyperon polarization for the first
    time in $J/\psi\to\Lambda\bar{\Lambda}$ reaction~\cite{BESIII:2018cnd} with both $\Lambda\to p\pi^-$  (and their conjugate) which allows the
    simultaneous determination of the $\Lambda$ and $\bar\Lambda$ decay asymmetries from the events.  With this novel method, a sensitive $CP$ symmetry, $A^{\Lambda}_{CP}$=$-0.006\pm0.012\pm0.007$~\cite{BESIII:2018cnd} , test in the strange baryon  sector is performed by directly comparing the asymmetry parameters for $\Lambda\to p\pi^-$ and $\bar\Lambda\to \bar p\pi^+$.
Most recently,  with the 10 billion $J/\psi$ events, BESIII updated this analysis and 
the polarization of $\Lambda$ has been observed, as demonstrated in Fig.~\ref{Lambda_mu}. Figure~\ref{Lambda_alpha} displays the results of the $\Lambda$ decay parameter obtained from various experiments. The $\alpha$ value determined in this study is in agreement with previous measurements from BESIII~\cite{BESIII:2018cnd} as well as the result extracted from the decay $J/\psi\to\Xi^-\bar\Xi^+$ reported by BESIII~\cite{BESIII:2021ypr}. However, it presents a deviation of 3.5$\sigma$ from the CLAS result.
In addition, the $A_{CP}$ superseded the previous measurement with  $-0.0025 \pm 0.0046 \pm 0.0012$~\cite{BESIII:2022qax}. 
The comparison to the previous works\cite{BES:2009zvb,Barnes:1996si, DM2:1988ppi}  and those from the cascade decays of hyperons\cite{BESIII:2021ypr, BESIII:2024nif,BESIII:2023jhj},  as illustrated in Table~\ref{table-lambda}, shows 
this is  the most precise measurement to  date. However, there is still much room to reach a precision level compatible with theoretical predictions at a level of $10^{-4}$~\cite{Donoghue:1986hh,Salone:2022lpt, He:2022xra}.

\begin{table}[!htb]
\caption{Summary of the results on $A^\Lambda_{CP}$  and $A_{CP}^\Sigma$ at BESIII from different analyses. The final BESIII value will combine these and forthcoming results.}
{\begin{tabular}{@{}|c|c|c|@{}}\hline
 Hyperons  & $A_{CP}$ ($\times10^{-3}$)  & Decay processes \\ \hline
 \multirow{5}{*}{$\Lambda$ decay} &$-2.5\pm 4.6\pm 1.2$ & $J/\psi\to\Lambda \bar \Lambda $ \cite{BESIII:2018cnd,BESIII:2022qax}\\
 ~ & $-37.0 \pm 117.0 \pm 90.0$& $J/\psi\to \Xi\bar \Xi$ \cite{BESIII:2021ypr}\\

 ~ & $-30.0 \pm 69.0 \pm 15.0$  & $J/\psi\to\Sigma^0\bar{\Sigma}^0 $ \cite{BESIII:2024nif} \\
 ~ & ${\color{white}-}6.9 \pm 5.8 \pm 1.8$    & $J/\psi\to\Xi^0\bar{\Xi}^0 $\cite{BESIII:2023drj} \\
~ & $ -4.0 \pm 7.0 ^{+3.0}_{-4.0}$ &  $J/\psi\to \Xi\bar \Xi$ \cite{BESIII:2023jhj} \\ \hline
 \multirow{3}{*}{$\Sigma$ decay}  & $-4.0\pm37.0\pm10.0$ & $J/\psi (\psi(3686))\to \Sigma^+\bar{\Sigma}^-$ \cite{BESIII:2020fqg}\\
 ~ & $-80.0\pm 52.0\pm 28.0$ &  $J/\psi (\psi(3686))\to \Sigma^+\bar{\Sigma}^-$ \cite{BESIII:2023sgt} \\
 ~ & $ 0.4 \pm 2.9 \pm 1.3$ & $J/\psi (\psi(3686))\to \Sigma^0\bar{\Sigma}^0$ \cite{BESIII:2024nif} \\


 \hline
\end{tabular}\label{table-lambda}}
\end{table}

\subsection{$J/\psi(\psi(3686))\rightarrow\Sigma\bar{\Sigma}$}



In case of  $\Sigma^{+}$,  a search for CP violation was performed in both $ J/\psi$ and $\psi(3686)$ decaying into $\Sigma^{+} \bar{\Sigma}^{-}$ pairs~\cite{BESIII:2020fqg}  with $1310.6\times10^{6}$ $J/\psi$  events and $448.1\times10^{6}$ $\psi(3686)$ events, respectively.
With a simultaneous fit for those two processes, the decay parameters  of $\Sigma^{+}$ and  $\bar{\Sigma}^{-}$  were determined and  the CP asymmetry $A^\Sigma_{\rm CP}$ was calculated to be  $-0.004\pm0.037\pm0.010$ for the first time, which was found to be consistent with CP conservation and was in agreement with the SM prediction of $\sim 3.6\times10^{-6}$ within present uncertainties~\cite{Tandean:2002vy}.
However,  an intriguing phenomena  was found that the directions of the $\Sigma^+/\bar{\Sigma}^-$ polarizations were observed to be opposite in $J/\psi$ and $\psi(3686)$ decays; this phenomenon, however, is not observed in $J/\psi \to \Xi
\bar{\Xi}$ and $\psi(3686) \to \Xi
\bar{\Xi}$ decays~\cite{BESIII:2022lsz, BESIII:2023lkg, BESIII:2023drj, BESIII:2021ypr}. Until now, there is no interpretation of these results, therefore, more experimental measurements of the hyperon polarization are highly desirable.

In addition, the quantum entangled $J/\psi \to
\Sigma^{+}\bar{\Sigma}^{-}$  pairs from
10 billion  $J/\psi$ events taken by the BESIII
detector are used to study the non-leptonic two-body weak decays of $\Sigma^+\to p \pi^0, \bar\Sigma^-\to \bar n \pi^- $ and $\Sigma^+\to
n \pi^+, \bar\Sigma^-\to \bar p \pi^0 $. The simultaneously determined decay parameters allow to investigate the $C\!P$ asymmetry and the corresponding $A^\Sigma_{C\!P}$ is measured to be $-0.080\pm0.052\pm0.028$. This is the first study to test
$C\!P$ symmetry in the hyperon to neutron decay, and the result is
consistent with $C\!P$-conservation.  while the theoretical predictions are at a level of $10^{-4}$ ~\cite{Donoghue:1986hh,Tandean:2002vy}


Meanwhile,  the large yield of quantum entangled $\Sigma^0 \bar{\Sigma}^0$ pairs selected from the 10 billion $J/\psi$ and  2.7 billion $\psi(3686)$ decays~\cite{BESIII:2024nif}  enables a pioneering test of
strong-$CP$ symmetry in the $\Sigma^0$ hyperon decays.  With the fitted parity-violating decay parameters of the decays $\Sigma^0 \to \Lambda \gamma$ 
    and $\bar{\Sigma}^0 \to \bar{\Lambda} \gamma$,   the strong-$CP$ symmetry is tested  by measuring the asymmetry 
    $A^{\Sigma}_{CP}= (0.4 \pm 2.9 \pm 1.3)\times 10^{-3}$. Furthermore, the spin polarizations of the $\Sigma^0$ hyperons with
opposite directions in the $J/\psi$ and $\psi(3686)$ decays are
observed for the first time. This phenomenon is also observed in the
case of the $\psi \to \Sigma^+ \bar{\Sigma}^-$ decays~\cite{BESIII:2020fqg}, 
but not in the $\Xi^{-(0)}$~\cite{BESIII:2023drj,BESIII:2023lkg, BESIII:2022lsz} hyperon pairs.

\subsection{$J/\psi(\psi(3686))\rightarrow\Xi\bar{\Xi}$}

\begin{table}[!htb]
\caption{Summary of the results on $\Delta\phi_{CP}$ and $A^\Xi_{CP}$ in $\Xi$ decays at BESIIII.}
{\begin{tabular}{@{}|c|c|c|@{}}\hline
$\Delta\phi_{CP}$ ($\times10^{-2}$ rad)  &  $A_{CP}^\Xi$ ($\times10^{-3}$) & Decay processes \\ \hline
$-0.48 \pm 1.37 \pm 0.29$  &  $6.0\pm13.4\pm5.6$ & $J/\psi\to \Xi^+\bar{\Xi}^-$ \cite{BESIII:2021ypr}\\
$-5.0 \pm 5.2 \pm 0.3$ &  $-150.0 \pm 510.0 \pm 100.0$ &  $\psi(3686)\to \Xi^+\bar{\Xi}^-$ \cite{BESIII:2022lsz} \\
$-0.01 \pm 0.69 \pm 0.09$ &   $-5.4 \pm 6.5 \pm 3.1$  & $J/\psi \to \Xi^0\bar{\Xi}^0$ \cite{BESIII:2023drj}\\
$ -0.3 \pm 0.8 ^{+0.3}_{-0.7}$  & $ -9.0 \pm 8.0 ^{+7.0}_{-2.0}$  & $J/\psi \to \Xi^-\bar{\Xi}^+$ \cite{BESIII:2023jhj}\\
$-7.9 \pm 8.2 \pm 1.0$   & $  -7.0$ $\pm$ 82.0 $\pm$ 25.0  & $\psi(3686) \to \Xi^0\bar{\Xi}^0$ \cite{BESIII:2023lkg}\\
 \hline
\end{tabular}\label{table-xi}}
\end{table}

A natural extension of these hyperon-polarization measurements is to investigate $\Xi^-$ decays in $J/\psi$ and $\psi(3686)$ decays, where the measurement of the polarization of the $\Xi^-$ hyperon is accessible via the process $\Xi^-\to{\Lambda\pi^-}$.
Using a sample of $1.31\times10^9$ $J/\psi$ events, this has recently been achieved by BESIII~\cite{BESIII:2021ypr} for  $J/\psi\to\Xi^-\overline\Xi{}^+$ and the ensuing sequential transitions $\Xi^-\to\Lambda\pi^-\to p\pi^-\pi^-$ and
$\overline\Xi{}^+\to\overline\Lambda\pi^+\to\overline p\pi^+\pi^+$. With the determination of decay parameters for $\Xi^-\to\Lambda\pi^-$ and  $\Lambda\to p\pi^-$,  as well as those for their anti-particles, three independent CP tests were performed.
The asymmetry $A^{\Xi}_{\rm CP}$ is measured for the first time and found to be $(6.0\pm13.4\pm5.6)\times10^{-3}$, 
while the corresponding SM prediction \cite{Tandean:2002vy} is $A_{\rm CP, SM}^{\Xi}=(-0.6\pm1.6)\times10^{-5}$.
Since this method enables a separation of strong and weak $\Xi\to\Lambda\pi$ decay amplitudes,  the weak phase difference, as one of the most precise tests of the CP symmetry for strange baryons, was determined to be $(\xi_1^{P} - \xi_1^{S}) =  (1.2\pm3.4\pm0.8)\times10^{-2}$~rad for the first time.   And the $\Delta \phi_\Xi$ is determined to be $ (-4.8\pm13.7\pm2.9)\times10^{-3}~{\rm rad}$, which is also in agreement with the CP conservation.
 The sequential $\Xi^-$ decays also provide an independent measurement of the $\Lambda$ decay parameters, which allows to test the CP symmetry and  the corresponding $A_{\rm CP}^{\Lambda}$ is determined to be $ (-3.7\pm11.7\pm9.0)\times10^{-3}$.
Most recently BESIII performed the analysis of   $J/\psi\to\Xi^-\overline\Xi{}^+$  with the intermediate states of  $\Lambda \to n\pi^0$ ~\cite{BESIII:2023jhj} and found the results, summarized in Table.~\ref{table-xi}, are consistent with the above measurements.

 In case of the $\Xi^0$, the  asymmetry parameters were measured using
entangled quantum $\Xi^0$-$\bar{\Xi}^0$ pairs~\cite{BESIII:2023jhj,BESIII:2023lkg}  and subsequent decays were investigated using 10 billion $J/\psi$ events and 448 million $\psi(3686)$ events , respectively.  
 The most precise values for $CP$ asymmetry observables of $\Xi^0$ decay are obtained to be $A^{\Xi}_{CP} = (-5.4
    \pm 6.5 \pm 3.1) \times 10^{-3}$ and $\Delta\phi^{\Xi}_{CP} =    (-0.1 \pm 6.9 \pm 0.9) \times 10^{-3}$~rad.    For the first time, the weak and strong phase differences are determined
    to be $\xi_{P}-\xi_{S} = (0.0 \pm 1.7 \pm 0.2) \times  10^{-2}$~rad and $\delta_{P}-\delta_{S} = (-1.3 \pm 1.7 \pm 0.4)   \times 10^{-2}$~rad, which are the most precise results for any weakly-decaying baryon.  
    
\subsection{$J/\psi\rightarrow\Lambda\bar{\Sigma}+c.c.$}

The decay of a psionic state $\psi$ into the pair of different hyperons $\psi\to\gamma^*\to \Lambda\bar\Sigma^0$ + c.c. violates the isospin symmetry and hence, assuming such a symmetry as exact, it must be purely electromagnetic, i.e., the coupling is mediated by a virtual photon: $\psi\to\gamma^*\to \Lambda\bar\Sigma^0$ + c.c., which was confirmed by the agreement between the expected coupling to the $J/\psi$ decay and the value extracted from cross section data in the electromagnetic continuum. As discussed in Ref.~\cite{Ferroli:2020xnv}, this feature enables to determine the the hyperon structure function.

\begin{figure}[!htbp]
	    \centering
	    \includegraphics[width=0.6\textwidth]{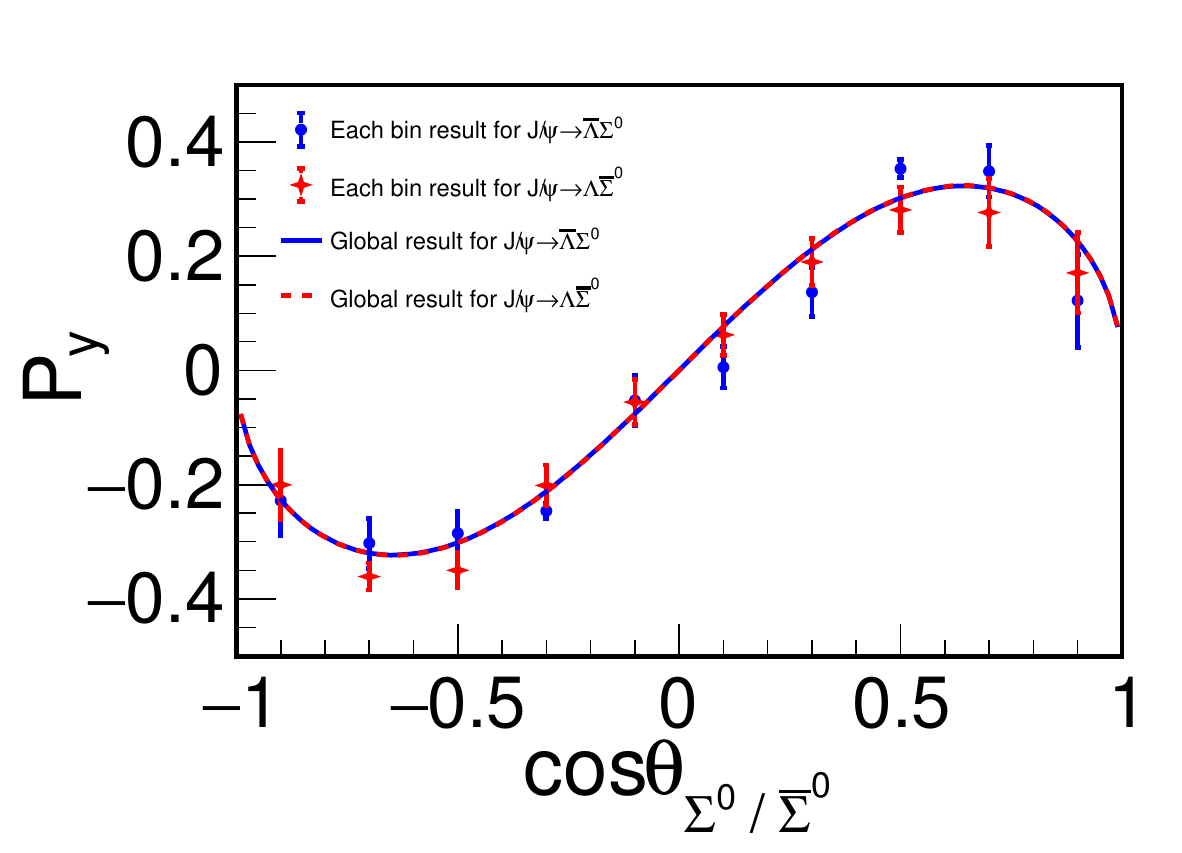}
	    \caption{Polarization in the $e^+e^-\to J/\psi\to\bar\Lambda\Sigma^0~(\Lambda\bar\Sigma^0)$ reaction. The curves show polarization $P_Y$ of the $\Sigma$ hyperon as a function of the emission angle.}
	    \label{LambdaSigma_py}
    \end{figure}

By exploiting quantum entangled pairs of $\Sigma^0~(\bar\Sigma^0)$ and $\bar\Lambda~(\Lambda)$, most recently BES\mbox{III} investigated the reaction $J/\psi \rightarrow\bar{\Lambda}\Sigma^{0}$~\cite{BESIII:2023cvk}.
 $\Lambda$ and $\bar{\Sigma}^0$ are not charge conjugates of each other which enables us to explore direct CP violation by comparison of polarizations from 
 both $J/\psi \rightarrow\Lambda\bar{\Sigma}^0$  and $J/\psi \rightarrow\bar\Lambda\Sigma^0$.   ${\Delta\Phi}_{\rm CP}$ is calculated to be $0.003\pm0.133\pm0.014$ rad, which is consistent with zero and indicates no evident direct CP violation. Moreover, the polarization of $\Sigma^0$ is presented in Fig.~\ref{LambdaSigma_py}.

Additionally,  since this is a purely electromagnetic process mediated by $\gamma^*\rightarrow J/\psi(loop)\rightarrow \gamma^*$, namely the hadronic vacuum polarization effect, which exhibits a notable enhancement attributed to the $J/\psi$ resonance. Therefore, BESIII probed the same vertex as the one-photon exchange process and attain the structure at the $J/\psi$ resonance. The result is essentially a precise snapshot of a $\bar\Lambda\Sigma^0$~($\Lambda\bar\Sigma^0$) pair in the making, encoded in the electromagentic form factor ratio and the phase. Their values are measured to be $R = 0.860\pm0.029({\rm stat.})\pm0.010({\rm syst.})$, $\Delta\Phi_1=(1.011\pm0.094({\rm stat.})\pm0.010({\rm syst.}))~\rm rad$ for $\bar\Lambda\Sigma^0$ and $\Delta\Phi_2=(2.128\pm0.094({\rm stat.})\pm0.010({\rm syst.}))~\rm rad$ for $\Lambda\bar\Sigma^0$, respectively.

\subsection{ $\psi(3686)\rightarrow \Omega^-\bar{\Omega}^+$\cite{BESIII:2020lkm}}

The discovery of the  $\Omega^-$ gave great credence to the SU(3) symmetry scheme of
particles and played an important role in establishment of the quark model in which it was expected to have a spin of 3/2.  
 However, this expectation has never been unambiguously confirmed by experiment before.
With the subdata sample of $448\times 10^6$ $\psi(3686)$ decays,  BESIII reported the  helicity amplitude analysis of $\psi(3686)\to \Omega^- \bar{\Omega}^+$
($\Omega^-\to K^-\Lambda$, $\bar{\Omega}^+\to K^+\bar{\Lambda}$,
$\Lambda\to p\pi^-$, $\bar{\Lambda}\to \bar{p}\pi^+$) with the selected 4035 $\pm$ 76 events. The decay
parameters of the subsequent decay $\Omega^-\to K^-\Lambda$
$(\bar{\Omega}^+\to K^+\bar{\Lambda})$ are measured for the first
time by a fit to the angular distribution of the complete decay chain, and found that the spin of the $\Omega^-$  favors $3/2$ as expected from the quark model. 
The helicity amplitudes of $\psi(3686)\to \Omega\bar{\Omega}$ and the decay parameter $\phi_{\Omega^{-}}$ of
$\Omega^{-}\to K^{-}\Lambda$, are also measured for the first
time.
The significance of $\phi_\Omega\ne0$ is 3.7 $\sigma$ and that for $\phi_\Omega\ne\pi$  is 1.5 $\sigma$ Thus, the D-wave-dominant (parity odd)decay is preferred,
which differs from the theoretical predictions of P-wave (parity even) dominance~\cite{Tandean:2004mv}. 



%
%
%

\section{Summary}

\begin{figure}[!htbp]
	    \centering
	    \includegraphics[width=0.8\textwidth]{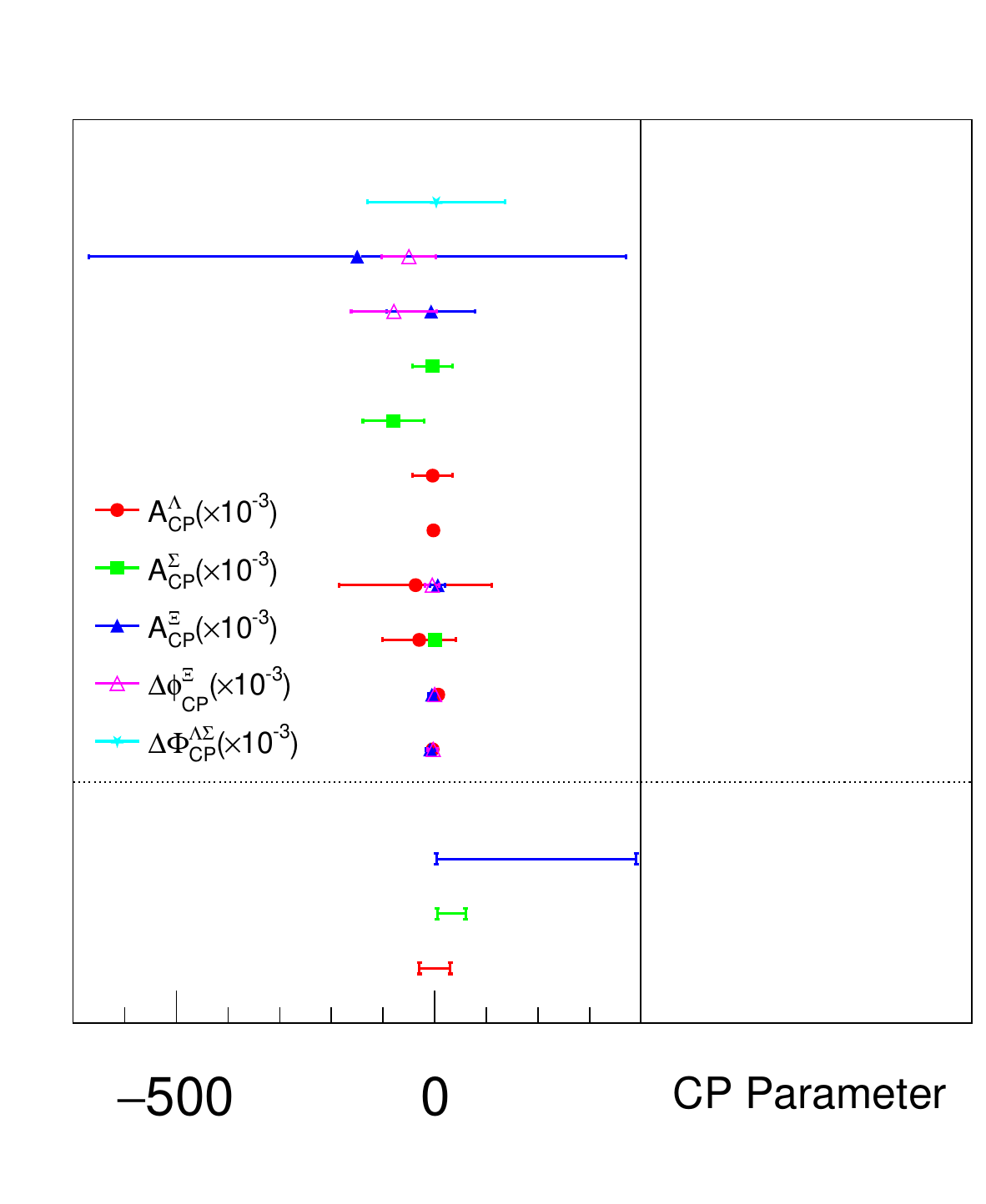}
	    \put(-125,402){\textcolor{red}{BESIII results}}
	    \put(-125,380){$J/\psi\to\Lambda\bar\Sigma$~\cite{BESIII:2023cvk}}
	    \put(-125,359){$\psi(3686)\to\Xi^+\bar\Xi^-$~\cite{BESIII:2022lsz}}
	    \put(-125,338){$\psi(3686)\to\Xi^0\bar\Xi^0$~\cite{BESIII:2023lkg}}
	    \put(-125,317){$J/\psi(\psi(3686))\to\Xi\bar\Xi$\cite{BESIII:2020fqg}}
	    \put(-125,296){$J/\psi(\psi(3686))\to\Xi\bar\Xi$\cite{BESIII:2023sgt}}
	    \put(-125,275){$J/\psi\to\Lambda\bar\Lambda$~\cite{BESIII:2018cnd}}
	    \put(-125,254){$J/\psi\to\Lambda\bar\Lambda$~\cite{BESIII:2022qax}}
	    \put(-125,233){$J/\psi\to\Xi\bar\Xi$~\cite{BESIII:2021ypr}}
	    \put(-125,212){$J/\psi\to\Sigma^0\bar\Sigma^0$~\cite{BESIII:2024nif}}
	    \put(-125,191){$J/\psi\to\Xi^0\bar\Xi^0$~\cite{BESIII:2023drj}}
	    \put(-125,170){$J/\psi\to\Xi\bar\Xi$~\cite{BESIII:2023jhj}}
	    \put(-125,149){\textcolor{red}{Theoretical prediction}}
	    \put(-125,128){$A^{\Sigma}_{\rm CP}(\times10^{-6})$~\cite{Tandean:2002vy}}
	    \put(-125,107){$A^{\Xi}_{\rm CP}(\times10^{-6})$~\cite{Salone:2022lpt}}
	    \put(-125,86){$A^{\Lambda}_{\rm CP}(\times10^{-6})$~\cite{Salone:2022lpt}}
	    \caption{ Comparison of the experimental measurements achieved at BESIII and theoretical expectations.
     }
	    \label{SumFig}
    \end{figure}
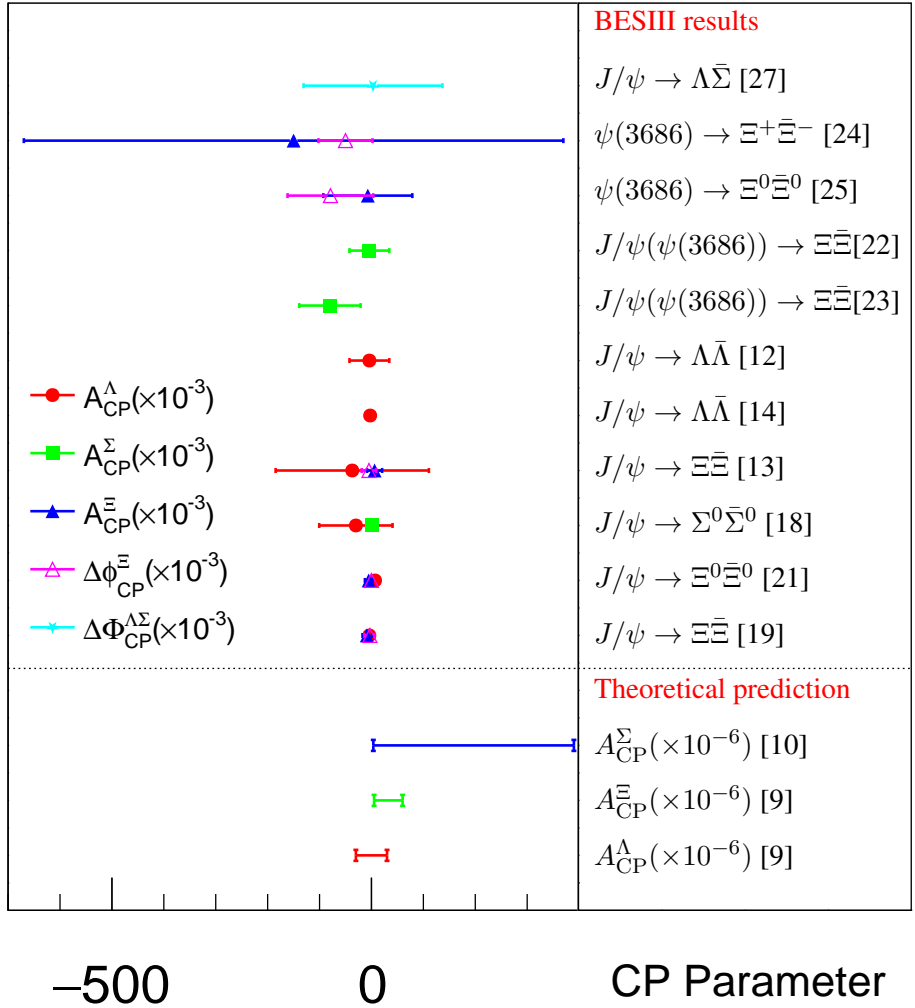

This review highlights the significant findings and implications of BESIII's research in advancing the field of CP violation through the investigation of charmonium decaying into entangled hyperon-anti-hyperon pairs selected from
the high statistics of  $J/\psi$ and $\psi(3686)$ data samples. 
The significant progresses achieved at BESIII, as illustrated in Fig.~\ref{SumFig}, primarily focusing on hyperons with a spin of 1/2, have proven to be of great interest in understanding the sources of CP violation and have attracted attention from both theoretical and experimental perspectives in hyperon physics. Currently, some of the results were obtained from the subdata samples of $J/\psi$ and $\psi(3686)$ events, which urgently need to be updated.
For hyperon-antihyperon pairs with spin of 3/2, research has only 
performed on $\Omega$ decays. In the near future, detailed investigations on other hyperon pairs with a spin of 3/2 will be carried out by the BESIII experiment, and important experimental results will be presented.

However, these achievements also highlight that the current statistics of hyperons at BESIII are still far from reaching the predictions of the SM.
In order to meet the  SM's expectation of $10^{-4}$,  a sample of $10^9$ reconstructed hyperons would be required, which is beyond the current capacity of the  BESIII experiment.  
 The ongoing upgrades to the BEPCII collider and BESIII detector aim to increase luminosity by a factor of three, enhance energy capabilities, and improve data collection efficiency, enabling more in-depth investigations into CP violation.
 Consequently, the BESIII experiment is expected to achieve higher sensitivity to direct CP violation in the near future. Despite these advancements, there will still be a disparity between the experimental results and the SM predictions.
 Fortunately,  discussions are underway in China regarding the construction of a Super Tau Charm Factory~\cite{Achasov:2023gey}  with a luminosity of  $10^{35}$cm$^{-2}$s$^{-1}$. This facility, compared to the BEPCII/BESIII, would increase the $J/\psi$ production rate by two orders of magnitude.  With advancements in detection technology, particularly the adoption of polarized beams~\cite{Salone:2022lpt}, and the innovative method developed by BESIII utilizing hyperon quantum correlations in decays to investigate CP violation,
 there is substantial potential to reach the range of CP violation effects predicted by the SM and potentially discover new phenomena with unprecedented statistics.

\section{Acknowledgments}

This work is supported in part by National Natural Science Foundation of China (NSFC) under Contracts No.  12225509; {Polish National Science Centre through the grant 2024/53/B/ST2/00975}.


\bibliography{bes3-mini-review-zwj_v2}
\end{document}